\documentclass{optica-article}

\journal{opticajournal} 

\articletype{Research Article}

\usepackage{lineno}

\usepackage[utf8]{inputenc}
\usepackage{multirow}
\usepackage{amsmath}
\usepackage{amsfonts}
\usepackage{graphicx}
\usepackage{array}
\usepackage{hyperref}
\usepackage{braket}
\usepackage{xcolor}
\usepackage{verbatim}
\usepackage{subfig}
\usepackage{ulem}
\usepackage{physics}
\usepackage{pdfpages}

\usepackage{lipsum}  

\begin{document}

\title{Optimized experimental optical tomography of quantum states of room-temperature alkali-metal vapor}

\author{Marek Kopciuch \authormark{1,2*}, Magdalena Smolis\authormark{2}, Adam Miranowicz\authormark{3}, Szymon Pustelny\authormark{2**}}

\address{\authormark{1}Doctoral School of Exact and Natural Sciences, Jagiellonian University, Faculty of Physics, Astronomy and Applied Computer Sciences, Łojasiewicza 11, 30-348 Krak\'{o}w, Poland\\
\authormark{2}Institute of Physics, Jagiellonian University in Krak\'ow, \L ojasiewicza 11, 30-348 Kraków, Poland\\
\authormark{3}Institute of Spintronics and Quantum Information, Faculty of Physics, Adam Mickiewicz University, 61-614 Pozna\'n, Poland\\}

\email{\authormark{*}marek.kopciuch@doctoral.uj.edu.pl} 
\email{\authormark{**}pustelny@uj.edu.pl}

\newcommand{\SP}[1]{\textbf{\color{red}{SP: #1}}}
\newcommand{\MK}[1]{\textbf{\color{green}{MK: #1}}}
\newcommand{\AM}[1]{\textbf{\color{blue}{AM: #1}}}

\renewcommand\vec{\mathbf}

\date{\today}

\begin{abstract}
We demonstrate a novel experimental technique for quantum-state tomography of the collective density matrix. It is based on measurements of the polarization of light, traversing the atomic vapor. To assess the technique's robustness against errors, experimental investigations are supported with numerical simulations. This not only allows to determine the fidelity of the reconstruction, but also to analyze the quality of the reconstruction for specific experimental parameters (light tuning and number of measurements). By utilizing the so-called conditional number, we demonstrate that the reconstruction can be optimized for a specific tuning of the system parameters, and further improvement is possible by selective repetition of the measurements. Our results underscore the potential high-fidelity quantum-state reconstruction while optimizing measurement resources.
\end{abstract}

\section{Introduction}

Quantum technology is built on the precise manipulation and reconstruction of quantum states. When dealing with single microscopic quantum objects, the reconstruction of states becomes challenging. This stems from a (often) destructive nature of the reconstruction and small amplitudes of recorded signals. To address these difficulties, some researchers have turned their focus towards studying ensembles of quantum objects, which display a collective quantum behavior \cite{Mouloudakis2022, Shaham2022, Happer1972, Schmidt1996, Hammer2004, Krauter2011, Katz2020, Jensen2011, Dabrowski2017, Mikhailov2008, Agha2010, Wasilewski2009, Heifetz2004, Dabrowski2016, Kopciuch2022Optical}. 

Atomic vapors serve as a prime example of a medium utilized for the engineering of collective quantum states. In their ultracold form, they allow for precise quantum control through light and other external fields, albeit implementation of the control requires complex experimental setups. On the other hand, room-temperature vapors can be studied using simpler apparatuses, but they simultaneously present challenges in terms of theoretical understanding \cite{Happer1972}. Despite these problems, however, the room-temperature atomic vapors were used to demonstrate various quantum-mechanical effects including coherent population trapping \cite{Schmidt1996}, spin squeezing \cite{Hammer2004, Krauter2011}, macroscopic entanglement \cite{Katz2020, Jensen2011}, spin waves \cite{Jensen2011, Dabrowski2017}, squeezed light generation \cite{Mikhailov2008, Agha2010, Hammer2004} and entanglement of light modes\cite{Wasilewski2009}. Rubidium vapor was also used to construct an on-demand quantum memory \cite{Heifetz2004, Dabrowski2016}. These experiments revived the interest in such media, while also necessitated the development of reliable quantum-state tomography (QST) methods.

In this work, we demonstrate the first experimental implementation of recently proposed QST method \cite{Kopciuch2022Optical}. The method enables the reconstruction of a collective density matrix of a room-temperature atomic vapor and is based on the illumination of the vapor with an off-resonant probing light and monitoring properties of the light after traversing a medium subjected to an external magnetic field. This enables to reconstruct a collective quantum state of $^{87}$Rb atoms residing in the $F=1$ ground state (qutrit). 

To evaluate the efficiency of the tomographic technique, we used the so-called conditional number \cite{Bogdanov2010, Adam2014, Bartkiewicz2016}. Previously, the parameter was used for a comprehensive comparison of tomographic methods of two polarization qubits~\cite{Adam2014}, NMR tomography of two $^{1}H$ spins-1/2 (two qubits)~\cite{Roy2010}, and a single nuclear spin-3/2 (a quartit) in a semiconductor quantum well~\cite{Adam2015}. We demonstrate that by an appropriate tuning of the probing light, the conditional number can be minimaized (corresponding to an optimized reconstruction) and as small as 2.25 can be achieved. We also discuss means of further improvement of the reconstruction efficiency by repeating specific measurements.

\section{Principles of the optical tomography}

We begin with a brief overview of the QST technique developed in Ref.~\cite{Kopciuch2022Optical}. This method relies on measuring the polarization rotation of linearly polarized probe light traversing a medium (e.g., room-temperature alkali metal atoms) subjected to a longitudinal magnetic field. We assume that the amplitude of the light is low, which allows us to describe its interaction with atoms using perturbation theory at the lowest order (linear interaction). At the same time, unlike previous approaches (see, e.g. Ref.~\cite{Deutsch2010Quantum, Lovecchio2015}), we do not assume a significant detuning of the light from the optical transition. This enables us to consider not only the vector contributions to a polarization rotation \cite{Hammer2010, Deutsch2010Quantum, Takahashi1999}, but also tensor one \cite{Colangelo2013}, and hence reconstruct the collective density matrix of the atoms. It is noteworthy that this reconstruction is achieved without full control over the system, as successive magnetic sublevels are equally splitted due to a weak magnetic field (under the conditions of the linear Zeeman effect) \cite{dalessandro2021}.

In Ref.~\cite{Kopciuch2022Optical}, the relation between time-dependent polarization rotation $\delta\alpha(t)$ and operators $\hat{\alpha}_{R,I}$ and $\hat{\beta}$ was introduced. The operators are associated with coherences and population difference of specific magnetic sublevels hence provide access to specific density-matrix elements. In this work, we employ a slightly modified version of that relationship, i.e.,
\begin{equation}
    \delta \alpha(t;\Delta) = \eta (\Delta) \left(  e^{-\gamma_1 t} \left[ \expval{\hat{\alpha}_{R}} \sin (2\Omega_L t)
    + \expval{\hat{\alpha}_{I}} \cos(2 \Omega_L t) \right] - \zeta(\Delta) e^{-\gamma_2 t} \expval{\hat{\beta}} \right), 
    \label{eq:rotationExp}
\end{equation}
where $\eta(\Delta)=\chi V_R(\Delta)$ and $\zeta(\Delta)=V_I(\Delta)/V_R(\Delta)$ are the so-called global and local scaling factors associated with real $V_R$ and imaginary $V_I$ parts of the Voigt profile, and $\chi$ is related to experimental parameters such as atomic density and transition frequency (for more details see the Supplemental Information -- SI). As shown in Eq.~\eqref{eq:rotationExp}, the time dependence of the polarization rotation is determined by the Larmor frequency $\Omega_L$ and the relaxation rates $\gamma_1$ and $\gamma_2$. 

Since a single measurement described by Eq.~\eqref{eq:rotationExp} allows us to extract only limited information about the system (specifically the population difference and coherence between magnetic sublevels with $\Delta m_F = 2$) it is necessary to expand the set of measured signals to obtain a more comprehensive information. To achieve this, we introduce a series of unitary operations known as control pulses, which systematically manipulate a given state in the Hilbert space. This provides the access to other density-matrix elements and hence offer a complete characterization of the system~\cite{Kopciuch2022Optical}. In turn, the reconstruction problem can be presented as
\begin{equation}
    \mathbb{O} \rho_{V} = \vec{b},
    \label{eq:linear_problem}
\end{equation}
where $\mathbb{O}$ represents the coefficient matrix determined by the set of observables, and $\rho_{V} =  \left[ \rho_{\bar{1}\bar{1}}^R, \rho_{\bar{1}0}^R, \rho_{\bar{1}0}^I, \ldots \right]^{T}$ (where $\rho_{mn}^R = {\rm Re}\lbrace \rho_{mn}\rbrace$, $\rho_{mn}^I = {\rm Im}\lbrace \rho_{mn}\rbrace$ and $\bar{1} = -1 $) is the vectorized form of a standard form density matrix $\rho$ with entrances $\rho_{ij}$ (see SI for more information), $\vec{b}$ is the observation vector and contains the values of the measured values of the observables. In a typical experimental scenario, the set of measurements given in Eq.~\eqref{eq:linear_problem} is often overdetermined, and it is advantageous to rescale it to a more suitable form
\begin{equation}
    \mathbb{C} \rho_{V} = \tilde{\vec{b}},
    \label{eq:linear_overdetermined}
\end{equation}
where $\mathbb{C} = \mathbb{O}^{\dagger}\mathbb{O}$ and $\tilde{\vec{b}} = \mathbb{O}^{\dagger} \vec{b}$. This rescaling enables the calculation of the density operator by simply inverting the aforementioned linear problem.

\section{Experimental details}

\subsection{Experimental setup}

\begin{figure}
    \centering
    \begin{tabular}{cc}
         \includegraphics[width=0.48\columnwidth]{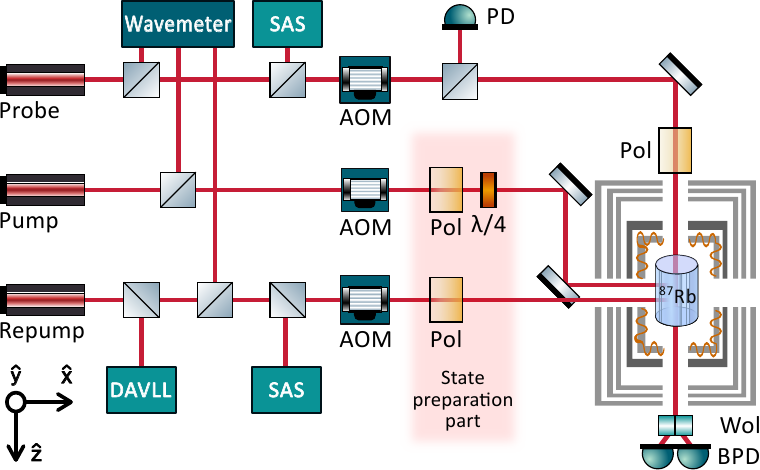} & 
         \includegraphics[width=0.48\columnwidth]{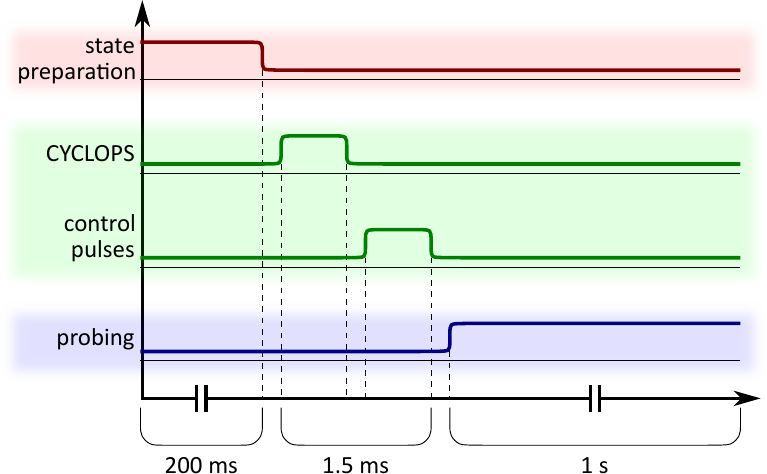}\\
         (a) & (b)
    \end{tabular}
    \caption{(a) Simplified scheme of the experimental setup used for the quantum-state generation and tomography. SAS -- saturation absorption spectroscopy, DAVLL -- dichroic atomic vapor laser lock, AOM -- acusto-optic modulator, PD -- photodiode, Pol -- Glan-Thompson polarizer, $\lambda/4$ -- quater-wave plate, $^{87}$Rb -- parafin-coated vapor cell filled with $^{87}$Rb, Wol -- Wollaston prism, BPD - balanced photodetector. (b) Experimental sequence used in our method. The initial state of atoms is prepared with the pump light turned on for about 200~ms (red trace). After the preparation period, we apply a sequence of magnetic pulses to modify the state of the atoms (green trace). Here, the CYCLOPS pulses (see Sec.~\ref{sec:CYCLOPS}) are first used and then the control pulses are implemented for a total time of about 1.5~ms. Finally, the probe light is turned on, alongside with the longitudinal magnetic field, for about 1~s (blue) and the polarization rotation signal is recorded.}
    \label{fig:experiment}
\end{figure}

The heart of our experimental system is a 3 cm diameter paraffin-coated spherical cell, containing an istopically enriched sample of $^{87}$Rb atoms. The cell is heated up to $50^{\circ}$C and is placed inside a cylindrical magnetic shield made of three layers of mumetal and a qubic innermost ferrite layer. Apart from the cell, the shield additionally contains a set of magnetic-field coils, which enables residual-field compensation and generation of magnetic-field pulses in the $\vec{x}$, $\vec{y}$, and $\vec{z}$ directions. Light used for the illumination of the rubidium atoms is provided by three diode lasers, where the pump and probe lasers are distributed-feedback lasers (DFBs), and the repump laser is the Fabry-Perot laser (ECDL). All lasers are independently tuned, and the repump laser wavelength is frequency-stabilized using a Dichroic Atomic Vapor Laser Lock (DAVLL) \cite{DAVLL2016}. The wavelengths of the other two lasers are passively maintained due to their inherent temporal stability. Performance of all lasers is monitored using a wavemeter, while the pump and probe lasers are additionally monitored through saturated absorption spectroscopy (SAS). The intensities of the laser beams are dynamically controlled by three acusto-optical modulators (AOMs). To generate a specific quantum state in the vapor, the pump-light polarization is set by polarizers (POLs) and quarter-wave plates ($\lambda/4$), while the repump light is linearly polarized orthogonal to the pump-light propagation direction (i.e., along $\vec{y}$-axis). To determine the local scaling factor (see discussion below), the intensity of the probe light is monitored and its $\vec{y}$ linear polarization prior to the shield is provided by a Glan-Thomson polarizer. Finally, the polarization rotation of the probe light is measured after the cell using a balanced polarimater consisting of a Wollastron prism (WOL) and a balanced photodetector (BPD). The schematic of the setup is shown in Fig.~\ref{fig:experiment}(a).

\subsection{Experimental sequence}

The experimental sequence utilized in our measurements is shown in Fig.~\ref{fig:experiment}(b). The sequence begins with a pumping period during which a specific quantum state is engineered. This stage typically consists of a 200 ms light pulse (optical pumping), which is applied simultaneously with the repumping that prevent the atoms from escaping into the dark ($F=2$) state, followed by a few short ($\approx100$~$\mu$s) magnetic-field pulses, enabling generation of a desired complex state. Subsequently, a series of magnetic-field pulses is used to mitigate technical problems (see Sec.~\ref{sec:CYCLOPS}), which is followed by a set of control pulses. Once the pulses are completed, a constant magnetic field along the $\vec{z}$-direction, ranging from 10-100~nT, is established. At the same time, a probe light beam, propagating along $\vec{z}$ with intensity of 1-10~$\mu$W/cm$^2$, is turned on. In order to improve the signal-to-noise ratio, the intensity of the probe light is modulated at a frequency of 200~kHz and the polarimeter signal is detected using a lock--in amplifier.

\subsection{Global and local scaling factor}

An important element of the reconstruction of the density matrix is the determination of the global scaling factor $\eta(\Delta)$ [see Eq.~\eqref{eq:rotationExp}]. This can be done by measuring the light absorption in an unpolarized vapor. Using the absorption relationship derived in the SI, the factor can be identified by comparing the absorption of the probe light, tuned to the same wavelength as that during the tomography measurements (i.e., blue-detuned from $f=1\rightarrow F=2$ by 50--400~MHz), with the absorption of far-detuned light (>15~GHz).
\begin{equation}
    \eta(\Delta) = \dfrac{27}{16} \left( \sqrt{\frac{U_2(\Delta)/U_1(\Delta)}{U_2(\infty)/U_1(\infty)}}- 1 \right),
    \label{eq:GSF}
\end{equation}
where $U_{1}$ is the voltage measured at the transimpedance photodetector placed in front of and $U_2$ after the medium (see Fig.~\ref{fig:experiment}(a) and the SI for more details) with $\Delta$ indicating the probe light tuned for QST and $\infty$ far-detuned light.

Experimental determination of the local scaling factor $\zeta(\Delta)$ [see Eq.~\eqref{eq:rotationExp}] presents a greater challenge. It requires preparation of an anisotropic, yet well-defined quantum state. In this work, we select ``stretched'' states that are generated along the $\vec{x}$- and $\vec{z}$-axes. The first state can be created by illuminating the atoms with a circularly polarized pump light propagating along the $\vec{x}$-axis. The preparation of the second state is more involved and requires the application of an additional magnetic-field pulse after the pumping, which rotates the atomic $\vec{x}$-polarization to the $\vec{z}$-direction (we have experimentally verified that this process did not introduce dephasing, as evidenced by the unchanged signal amplitude for a many-$\pi$ pulse). Employing this procedure allows us to mitigate potential systematic errors arising from varying polarization levels achieved with the pump light propagating along different directions, while simultaneously simplifying the experimental setup. The formulas for the light polarization rotation corresponding to these two states are (see the SI for more details)
\begin{subequations}
    \begin{eqnarray}
        \delta \alpha^{(z)} (t;\Delta) &=& -\dfrac{5(1-\epsilon)}{24} \eta(\Delta) \zeta (\Delta) e^{-\gamma_2 t},\\
        \delta \alpha^{(x)} (t;\Delta) &=& -\dfrac{(1-\epsilon)}{48} \eta(\Delta) e^{-\gamma_1 t} \cos (2 \Omega_L t),
    \end{eqnarray}
    \label{eq:LSF_observables}
\end{subequations}
where $\epsilon$ is the remaining isotropic part of state. This allows one to calculate the local scaling factor
\begin{equation}
    \zeta(\Delta) = \dfrac{1}{10} \dfrac{\delta\alpha^{(z)}(0; \Delta)}{\delta\alpha^{(x)}(0; \Delta)}.
    \label{eq:LSF}
\end{equation}

\subsection{CYCLOPS-like measurement \label{sec:CYCLOPS}}

Equation~\eqref{eq:rotationExp} shows that our reconstruction method is sensitive to the initial phase of the measured signal. As uncontrollable phase delays are present in every experiment, the identification of the quadrature components of the signal becomes difficult. To address this issue, we adapt the CYCLically Ordered Phase Sequence (\mbox{CYCLOPS}) method, commonly utilized in nuclear magnetic resonance experiments \cite{Freeman1987, Bonk2005}. In our approach, we leverage the fact that the $\pi$-rotation of the state around the $\vec{y}$-axis leads to a sign reversal of $\expval{\hat{\alpha}_{I}}$ and $\expval{\hat{\beta}}$ (for more information, see the SI). At the same time, by applying the pulse rotating the state by $\pi/2$ around the $\vec{z}$-axis and next the pulse rotating the state around the $\vec{y}$-axis by $\pi$ (see the SI) the signs of the $\expval{\hat{\alpha}_{R}}$ and $\expval{\hat{\beta}}$ are reversed. By subtracting these two transformed states from the initial signal, we obtain
\begin{subequations}
    \begin{eqnarray}
        \left( \delta \alpha - \delta \alpha^{(Y)} \right) (t; \Delta) = 2 \eta (\Delta) \left[ -  \zeta(\Delta) e^{\gamma_2 t} \expval{\hat{\beta}}+e^{-\gamma_1 t}  \expval{\hat{\alpha}_{I}} \cos(2 \Omega_L t + \varphi) \right],
        \label{eq:rotationExp_cyc_Y}\\
        \left( \delta \alpha - \delta \alpha^{(ZY)} \right) (t; \Delta) = 2 \eta (\Delta) \left[ -  \zeta(\Delta) e^{\gamma_2 t} \expval{\hat{\beta}} + e^{-\gamma_1 t}  \expval{\hat{\alpha}_{R}} \sin(2 \Omega_L t + \varphi) \right],
        \label{eq:rotationExp_cyc_ZY}
    \end{eqnarray}
    \label{eq:rotationExp_cyc}
\end{subequations}
where $\varphi$ is an unknown phase shift originating from the experimental apparatus. In our CYCLOPS-like measurements, the problem of unknown phase is alleviated, as the final signals [see Eqs.~\eqref{eq:rotationExp_cyc}] depend only on one quadrature (via either sine or cosine time dependence) and, thus, $\varphi$ becomes insignificant. The procedure also allows us to remove systematic shifts of the signals associated with the imbalance of the polarimeter (for more details, see the SI).

\section{Reconstruction of states}

To perform QST, we conducted the above-described nine measurements, consisting of three sets of CYCLOPS-like pulses for each of three control pulses. To ensure the self-consistency of our reconstruction procedure, we simultaneously fit all of the polarization-rotation signals with shared parameters such as the global phase, relaxation rates, and oscillation frequency. The fitting values are then used to determine the observables and reconstruct the qutrit density-matrix elements using the linear inversion method given in Eq.~\eqref{eq:linear_overdetermined}. However, as this method does not guarantee the reconstructed matrices to be positive semidefinite, we utilize the maximum likelihood method with the Euclidean norm \cite{Adam2015, Paris2004} to find the closest physical realization of the reconstructed matrix.

To validate our tomography technique, we compare the reconstructed density matrices with numerical simulations of the state obtained during the pumping stage. For the simulations, we assume the interaction of an appropriately polarized light with a Doppler-broadened medium consisting of atoms of the energy-level structure similar to that of the $D_1$ line in $^{87}$Rb. As in the real experiment, we assume that there are two distinct regions between which the atoms can freely move. In the first region, the atoms evolve in a homogeneous magnetic field and relax to thermal equilibrium due to the collisions with vapor-cell walls and between one another. This corresponds to the atoms residing outside of the light beams. In the second region, the atoms still interact with the magnetic field but also with the pump and repump light. Moreover, we neglect the wall relaxation in this region. The latter region corresponds to the atoms inside the light beams. All parameters used in the simulations match the parameters of our experimental setup.
\begin{figure}
    \centering
    \begin{tabular}{ccc}
         \includegraphics[width = 0.4\columnwidth]{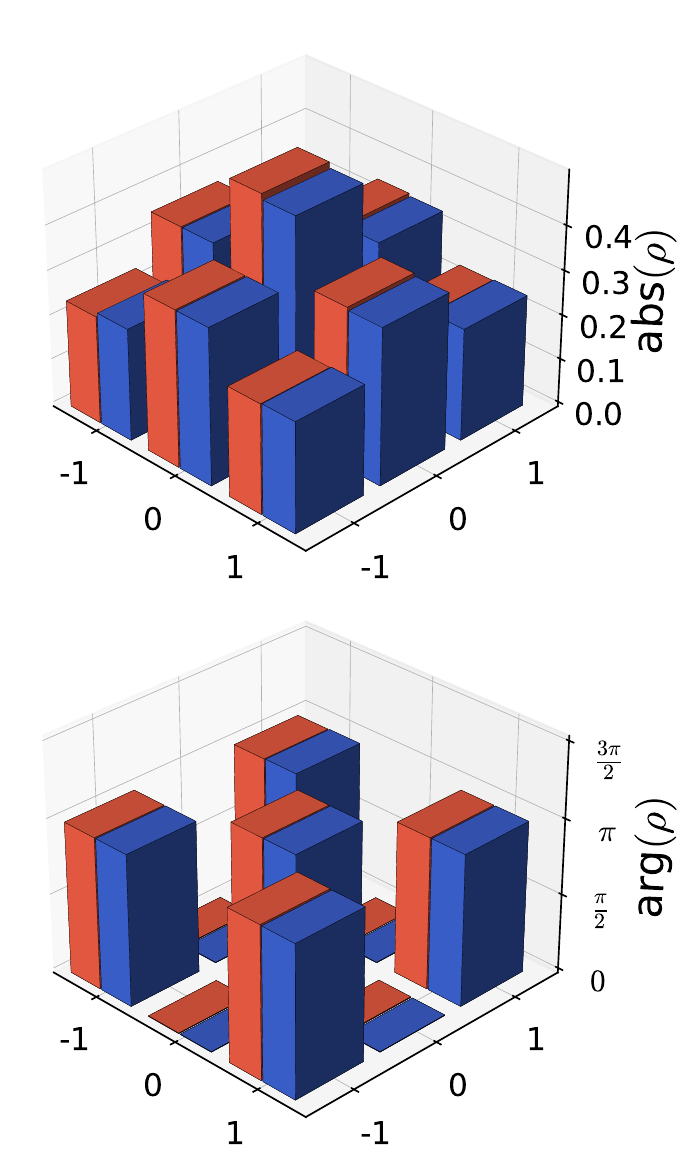}& \hspace{0.01\textwidth} &
         \includegraphics[width = 0.4\columnwidth]{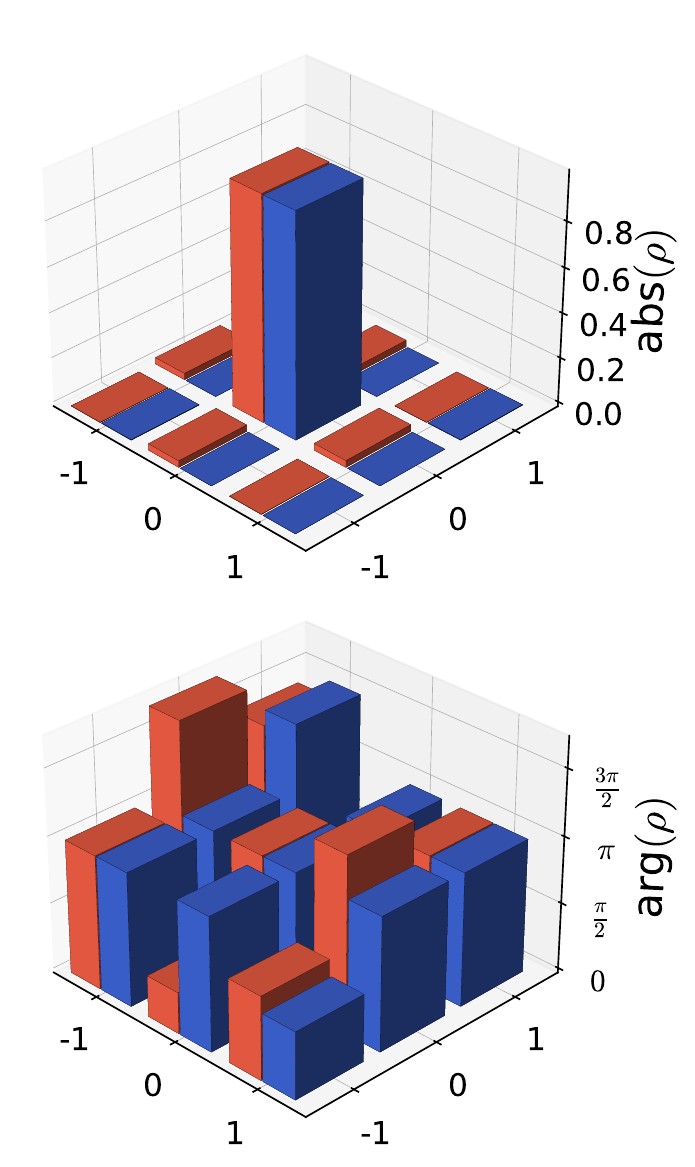}\\
         (a) & & (b)
    \end{tabular}
    \caption{Comparison of two experimentally reconstructed density matrix elements (blue bars), with simulated ones (red bars), including their amplitude (upper plots) and phase (lower plots). Here we chose simple pumping schemes with (a) circularly polarized and (b) linearly $\vec{z}$--polarized pumps propagating along $\vec{x}$--axis. The fidelity achieved between the experimental results and simulations in both cases exceeds 0.99.}
    \label{fig:reconstruction}
\end{figure}

As representative examples for our reconstruction, we consider two states that can be easily generated experimentally and simulated theoretically. The first state can be pumped with a strong, circularly polarized pumping light, propagating along the $\vec{x}$-axis [Fig.~\ref{fig:reconstruction}(a)]. The state has a nonuniform population distribution and its all coherences are nonzero. This allows us to demonstrate that our method can reconstruct not only different coherences but also determine their amplitudes and phases with a high accuracy. The results of the experimental reconstruction and simulations are presented in Fig.~\ref{fig:reconstruction}(a). As seen, the results are in a very good agreement revealing a reconstruction fidelity of 0.995. As the second example, we considered a state pumped with the $\pi$-polarized light, propagating along the $\vec{x}$-axis. In the ideal case (without experimental artefacts), this scheme leads to the total depletion of the $m_F=\pm 1$ states and no coherences between any sublevels. As shown in Fig.~\ref{fig:reconstruction}(b), our measurements demonstrate a good agreement with numerical simulations, demonstrating the fidelity of 0.998. Nonetheless one can notice that a very small amplitude of the coherences can lead to the deterioration of the phase reconstruction. The very high quality of the reconstruction of these two representative states demonstrates the usefulness of our QST technique.

\section{Conditioning and optimization of quantum state tomography}

\subsection{Condition number in linear inversion \label{sec:Conditional}}

As mentioned above, the condition number $\kappa$ is a useful parameter to evaluate the reliability of a QST method [see Eq.~\eqref{eq:linear_overdetermined}]. Specifically, to quantify the ability to tolerate errors or sensitivity to the errors, we use the condition number of a (nonsingular) matrix $\mathbb{C}$, which, assuming the spectral norm $\|\dots\|_2$, can be defined as  \cite{AtkinsonBook, HighamBook, GolubBook}
\begin{equation}
\kappa(\mathbb{C}) = \Vert \mathbb{C} \Vert_2\; \Vert \mathbb{C}^{-1}\Vert_2  =\max[{\rm svd}(\mathbb{C})]\max[{\rm svd}(\mathbb{C}^{-1})] =\dfrac{\max[{\rm svd} (\mathbb{C})]}{\min[{\rm svd}(\mathbb{C})]} \ge 1, \label{ConditionNoDef}
\end{equation}
where ${\rm svd}(\mathbb{C})$ denotes the singular values of $\mathbb{C}$. The significance of this error-robustness parameter explains well the Gastinel-Kahan theorem~\cite{HighamBook}, which states that a relative distance of a nonsingular square matrix $\mathbb{C}$ from the set of singular matrices corresponds to the inverse of a condition number. Utilizing the error $\delta\tilde{\vec{b}}$ in the observation vector $\tilde{\vec{b}}$ and the condition number $\kappa(\mathbb{C)}$, one can estimate the error $\delta \rho_{V}$ in the reconstructed density matrix $\rho_{V}$ from the so-called Atkinson inequalities~\cite{AtkinsonBook}
\begin{equation}
 \dfrac{1}{\kappa(\mathbb{C})} \dfrac{\Vert\delta \tilde{\vec{b}} \Vert}{\Vert\tilde{\vec{b}}\Vert} \le  \dfrac{\Vert\delta \rho_{V}\Vert}{\Vert\rho_{V}\Vert} \le \kappa(\mathbb{C}) \dfrac{\Vert\delta \tilde{\vec{b}}\Vert}{\Vert\tilde{\vec{b}}\Vert}.
\label{eq:Atkinson1}
\end{equation}
When the condition number approaches 1, it becomes apparent that small relative variations in the observation vector $\tilde{\vec{b}}$ result in correspondingly small relative changes in the reconstructed state $\rho_{V}$. In order to account for errors $\delta \mathbb{C}$ present in the coefficient matrix $\mathbb{C}$, these inequalities can be expanded according to the formulation derived in Ref.~\cite{AtkinsonBook}, giving rise to the expression
\begin{equation}
 \dfrac{\Vert\delta \rho_{V}\Vert}{\Vert\rho_{V}\Vert}
 \le
 \dfrac{\kappa (\mathbb{C})}{1- \kappa (\mathbb{C})
 \Vert\delta\mathbb{C}\Vert/\Vert\mathbb{C}\Vert}
 \left[\dfrac{\Vert\delta \tilde{\vec{b}}\Vert}{\Vert\tilde{\vec{b}}\Vert}+
 \dfrac{\Vert\delta\mathbb{C}\Vert}{\Vert\mathbb{C}\Vert}\right].
\label{eq:Atkinson2}
\end{equation}
By referring to the inequalities in Eqs.~(\ref{eq:Atkinson1}) and~(\ref{eq:Atkinson2}), we can infer that the quality of a QST method, in terms of its error sensitivity or robustness, can be assessed through its condition number $\kappa(\mathbb{C})$, which characterizes the degree to which small (large) changes in the observation vector $\tilde{\vec{b}}$ lead to relatively small (large) changes in the reconstructed state $\rho_{V}$. Thus, if $\kappa(\mathbb{C})$ is small (large), the QST method is well-conditioned (ill-conditioned), indicating the robustness (sensitivity) of the method to errors in the observation vector $\tilde{\vec{b}}$. In the case of ill-conditioned QST, even slight errors in $\tilde{\vec{b}}$ can cause significant errors in the reconstructed $\rho_{V}$. In short, the smaller the condition number the stronger robustness of a given linear-inversion-based QST method against errors. Thus, one can refer to an optimal method in this respect if $\kappa(\mathbb{C})=1$. Numerical examples of ill-conditioned QST problems can be found in Refs.~\cite{AtkinsonBook, Adam2014}.

\subsection{Optimization via probe light tuning}

In order to optimize a QST process, it is desired to make the coefficient matrix $\mathbb{C}$ more isotropic, which means that each measurement brings an equal amount of information about the system. A simple example of such an optimized problem is when each measurement brings information about only a specific density-matrix element, with all measurements having the same weight \cite{Adam2014, Bartkiewicz2016}. In this case, the coefficient matrix $\mathbb{C}$ is proportional to identity. Even though such optimization is intuitive, it is often unpractical, as experimental transformations required to achieve a desired scheme are very complex. Instead, here we propose a scheme, where a single experimental parameter is adjusted. In our case, this parameter is the probing light detuning, which, incorporated in Eq.~\eqref{eq:rotationExp} through $\zeta(\Delta)$, makes one of the observable detuning dependent. 

\begin{figure}
    \centering
    \begin{tabular}{cc}
         \includegraphics[width = 0.45\columnwidth]{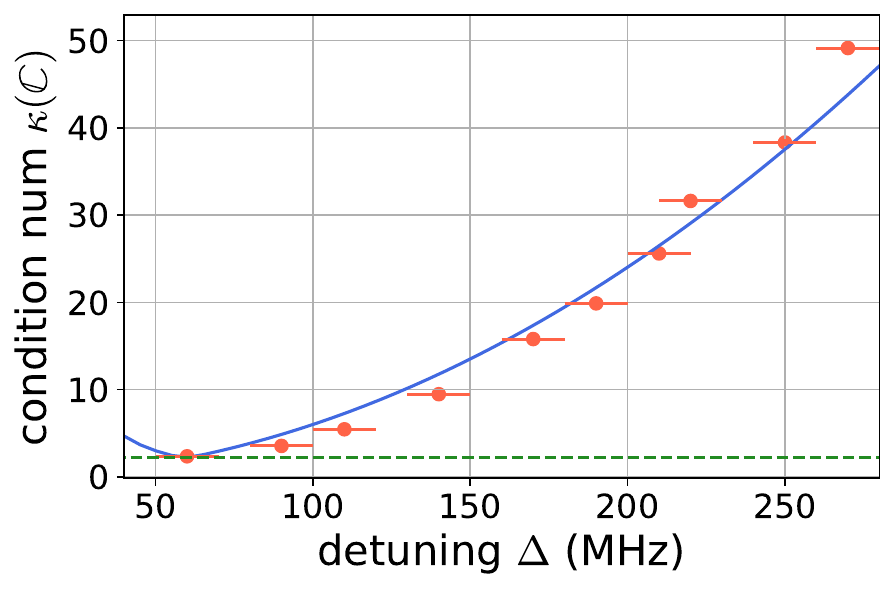} &
         \includegraphics[width = 0.45\columnwidth]{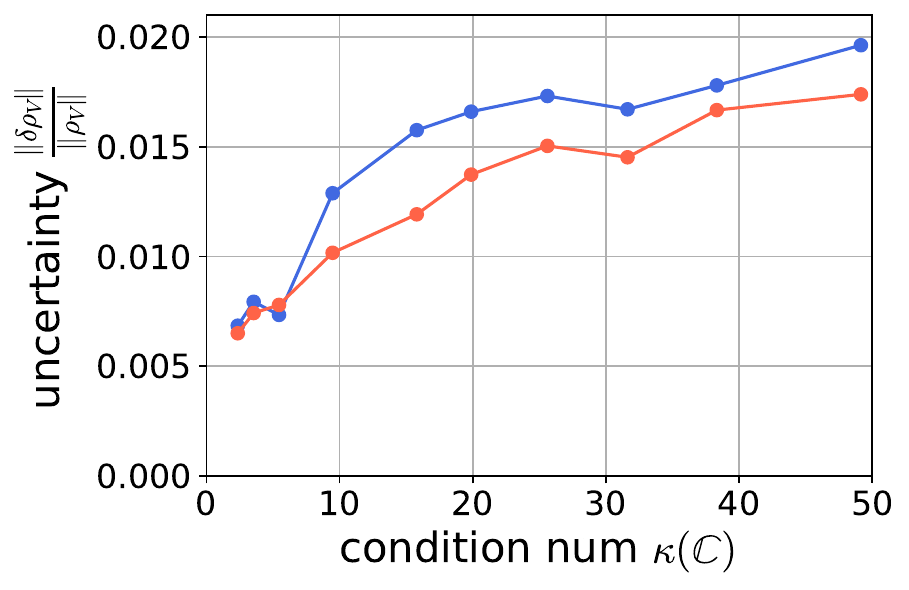}\\
         (a) & (b)
    \end{tabular}
    \caption{(a) Condition number of a state measured versus the probing-light detuning from the center of the Doppler-broadened $f=1\rightarrow F=2$ transition. The red points indicate the values calculated based on experimental measurements (horizontal uncertainty comes from the uncertainty of detuning, the evaluated vertical errors are small hence not visible), while the blue line shows the theoretical dependence calculated from the absorption measurements. Green dashed line indicates the smallest $\kappa(\mathbb{C}) = 2.25$ achievable using this approach. (b) Relative uncertainty of the linear inversion [see Eq.~\eqref{eq:Atkinson1}] as a function of the condition number $\kappa(\mathbb{C})$. Red points correspond to the reconstruction of the state pumped with circularly polarized light propagated along the $\vec{x}$-axis [see Fig.~\ref{fig:reconstruction}(a)] and blue points correspond to the reconstruction of the state generated with linearly $\vec{z}$-polarized pump light, propagating along the $\vec{x}$-axis [see Fig.~\ref{fig:reconstruction}(b)]. The uncertainty of the condition number is significantly smaller than the size of the data points and solid lines are added for clarity.}
    \label{fig:condition}
\end{figure}

It is important to note that our method does not guarantee an optimal tomography process, $\kappa(\mathbb{C}) = 1$. Therefore, to explore the limit of the method, we calculate the eigenvalues of coefficient matrix with $\zeta(\Delta)$ as a free parameter. In our case, the eigenvalues of $\mathbb{C}$ can be analytically calculated, taking the values: $\left\{\dfrac{1}{100},\dfrac{1}{150},\dfrac{1}{225},\dfrac{1}{225},\dfrac{1}{225},\dfrac{\zeta^2}{18},\dfrac{\zeta^2}{9},\dfrac{\zeta^2}{9}\right\}$. From this, we obtain the dependence of $\kappa(\mathbb{C})$ on $\zeta(\Delta)$ [Fig.~\ref{fig:condition}(a)] and a minimal possible conditional number of 2.25 is determined.

To further illustrate the effect of the probing-light detuning on the reconstruction uncerainty and, hence, demonstrate the potential of this approach, we perform a series of reconstructions of a state generated under the same conditions but reconstructed using different probing-light detunings. In our experiment, the detuning is changed from 50 to 270~MHz. The results of these investigations are shown in Fig.~\ref{fig:condition}(b). They demonstrate that the reliability of the reconstruction deteriorates with the detuning and it achieves the minimum in the vicinity of the center of the Doppler-broadened $f=1\rightarrow F=2$ transition. This agrees with our theoretical prediction of the condition-number detuning dependence, which we calculate assuming that $\zeta(\Delta)=V_I(\Delta)/V_R(\Delta)$.

\subsubsection{Conditional number versus the number of measurements}

\begin{figure}
    \centering
    \includegraphics[width=0.5\columnwidth]{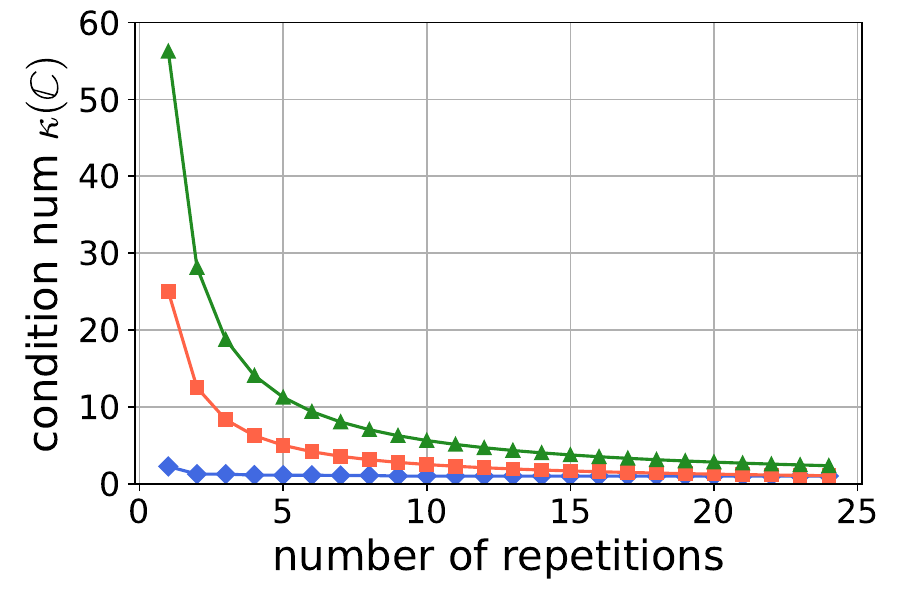}
    \caption{Minimum achievable condition number versus a number of repetitions of specific measurements. Three sets of points correspond to the condition number at different detunings: 300~MHz (green triangles), 200~MHz (red squares), and 60~MHz (blue diamonds). Solid line used for eye guidance.}
    \label{fig:cond_vs_num}
\end{figure}

The repetition of specific measurements offers a straightforward and versatile method for optimizing the relative weights of the observables used in the state-reconstruction procedure. This approach allows for achieving an arbitrarily small condition number, making it particularly valuable when the previous method is infeasible or when the condition number is desired to be smaller than the detuning-optimized bound (e.g., 2.25). However, it should be noted that this technique is associated with a potential drawback; the number of repetitions required to attain $\kappa(\mathbb{C})=1$ is typically substantial, especially when dealing with initially high condition numbers, as illustrated in Fig.~\ref{fig:cond_vs_num}. 

\section{Conclusions}

In this study, we presented the first experimental implementation of a quantum-state tomography technique, which was originally proposed in Ref.~\cite{Kopciuch2022Optical}. The technique enabled us the successful reconstruction of collective quantum states of a qutrit in room-temperature rubidium vapor at the $f=1$ ground state with a fidelity of 0.99. To overcome experimental challenges of the reconstruction, we adapted the CYCLOPS technique, which allowed us to achieve reliable reconstruction by mitigating a problem of unknown phase delays present in measured signals. Additionally, we presented a comprehensive analysis of the technique by introducing the conditional number, which quantifies the reliability of the reconstruction. This parameter was investigated versus different experimental factors, including tuning of the probing light used for the reconstruction. We demonstrated that by appropriate tuning of the light, the conditional numbers as low as 2.25 can be achieved (where the conditional number of 1 refers to ideal reconstruction). We also demonstrated that further improvement of the reconstruction (lowering the conditional number) can be achieved by the repetition of the specific measurements.

The successful implementation of the presented QST technique opens up avenues for measuring a range of fundamental properties of qutrits. In future, we plan to focus on exploring different measures of nonclassicality and establishing their ordering for various classes of quantum states. We also plan on a further development of the technique to demonstrate quantum-process tomography, expanding the method capabilities in the characterization of quantum operations and transformations. Finally, the ability to accurately reconstruct the quantum states of atomic ensembles allows for experimental optimization of generation of metrologically appealing quantum states. This is the research direction that we currently pursuit in our work.

\section{Acknowledgements} 

The authors would like to thank Arash D. Fard for his help in experimental measurements. The work was supported by the National Science Centre, Poland within the SONATA BIS programme (Grant No. 2019/34/E/ST2/00440). MK would like to acknowledge support from the Excellence Initiative -- Research University of the Jagiellonian University in Krak\'ow. A.M. is supported by the Polish National Science Centre (NCN) under the Maestro Grant No. DEC-2019/34/A/ST2/00081.

\bibliography{bibliography}



\section*{Supplementary material}

# Optimized experimental optical tomography of quantum states of room-temperature alkali-metal vapor: supplementary information

**Marek Kopciuch** [1,2*], **Magdalena Smolis** [2], **Adam Miranowicz** [3], **Szymon Pustelny** [2**]

[1] *Doctoral School of Exact and Natural Sciences, Jagiellonian University, Faculty of Physics, Astronomy and Applied Computer Sciences, Łojasiewicza 11, 30-348 Kraków, Poland*
[2] *Institute of Physics, Jagiellonian University in Kraków, Łojasiewicza 11, 30-348 Kraków, Poland*
[3] *Institute of Spintronics and Quantum Information, Faculty of Physics, Adam Mickiewicz University, 61-614 Poznań, Poland*
[*] *marek.kopciuch@doctoral.uj.edu.pl*
[**] *pustelny@uj.edu.pl*

## 1. Principles of the tomography method

In Ref. [1], the relation between the time- and detuning-dependent polarization rotation $\delta\alpha(t;\Delta)$ and the operators $\hat{\alpha}_{R,I}$ and $\hat{\beta}$, associated with the coherences and populations difference of specific magnetic sublevels, was introduced

$$\delta\alpha(t;\Delta) = \chi e^{-\gamma t} \left[ \langle \hat{\alpha}_R \rangle V_R(\Delta) \sin(2\Omega_L t) + \langle \hat{\alpha}_I \rangle V_R(\Delta) \cos(2\Omega_L t) - \langle \hat{\beta} \rangle V_I(\Delta) \right], \quad (1)$$

where $V_R(\Delta)$ and $V_I(\Delta)$ are the real and imaginary parts of the Voigt profile, respectively, and $\chi$ is related to experimental parameters such as an atomic density and transition frequency. In order to derive Eq. (1), isotropic relaxation of the atomic state was assumed. This may not be the case in a real experiment, where various experimental factors, e.g., magnetic-field gradients, may introduce additional dephasing of coherences. In turn, off-diagonal elements of the density matrix relax faster than the diagonal ones. As the observables $\hat{\alpha}_{R,I}$ depend exclusively on the coherences, while $\hat{\beta}$ on the population imbalance, to account for the fact, we introduce two separate relaxation rates, slightly modifying Eq. (1)

$$\delta\alpha(t;\Delta) = \eta(\Delta) \left( e^{-\gamma_1 t} \left[ \langle \hat{\alpha}_R \rangle \sin(2\Omega_L t) + \langle \hat{\alpha}_I \rangle \cos(2\Omega_L t) \right] - \zeta(\Delta) e^{-\gamma_2 t} \langle \hat{\beta} \rangle \right), \quad (2)$$

where $\eta(\Delta) = \chi V_R(\Delta)$ and $\zeta(\Delta) = V_I(\Delta)/V_R(\Delta)$ are the so-called global and local scaling factors, respectively.

The expectation values $\langle \hat{\alpha}_{R,I} \rangle$ and $\langle \hat{\beta} \rangle$ depend on the population deference and coherences between the states of the magnetic quantum-number difference $\Delta m_F = 2$. Thus, the operators do not allow to reconstruct an entire density matrix of the system. In order to access the remaining density-matrix elements (e.g., the coherences with the $\Delta m_F = 1$), one needs to transform the measured observables, so that the overall set of measurements defines a positive operator-valued measure (POVM) [2]. This can be achieved by introduction the unitary–evolution operators $U_i$, commonly called the control pulses, which modify the observable $\hat{O}$ as follows

$$\text{Tr}\left[\hat{O} U_i \rho U_i^\dagger\right] = \left\langle U_i^\dagger \hat{O} U_i \right\rangle = \left\langle \hat{O}' \right\rangle, \quad (3)$$

where $\hat{O}$ can be either $\hat{\alpha}_{R,I}$ or $\hat{\beta}$.

With the set of the observables $\hat{O}_i$, one can present the reconstruction problem in a vectorised form,

$$\mathbb{O} \rho_V = \mathbf{b}, \quad (4)$$

where $\mathbb{O}$ is the coefficient matrix determined by the set of observables and $\rho_V = \left[\rho^R_{\bar{1}\bar{1}}, \rho^R_{\bar{1}0}, \rho^I_{\bar{1}0}, \ldots\right]^T$ is the vectorise form of the density matrix with only real entrances and superscripts $R$ and $I$

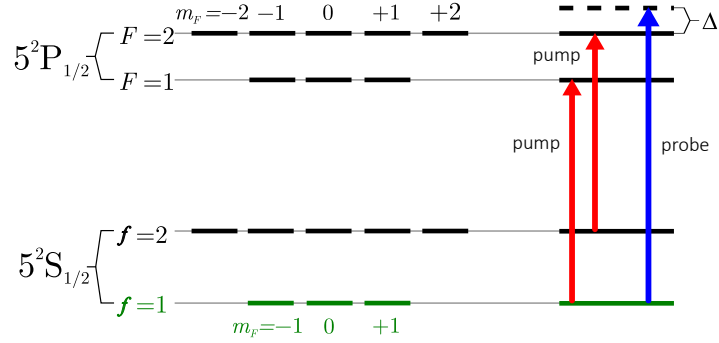

Fig. 1. Energy-level diagram corresponding to the $^{87}$Rb D$_1$ line (presented energy gaps are not in scale), with measured qutrit marked with green. Additionally, tunings of all laser beams are presented, red lines correspond to the pumping and repumping beams used for state preparation, and the blue line shows the probing beam which is blue detuned with detuning $\Delta$.

indicate real and imaginary parts of specific density-matrix elements. The vector **b** is called the observation vector and it contains values of all measured observables.

In a typical experimental scenario, the set of equations given in Eq. (4) is overcomplete and it is convenient to rescale it to

$$\mathbb{C}\rho_V = \tilde{\mathbf{b}}, \qquad (5)$$

where $\mathbb{C} = \mathbb{O}^\dagger \mathbb{O}$ and $\tilde{\mathbf{b}} = \mathbb{O}^\dagger \mathbf{b}$. This allows one to calculate the density operator by the inversion of the $\mathbb{C}$, i.e.,

$$\rho_V = \mathbb{C}^{-1}\tilde{\mathbf{b}}. \qquad (6)$$

It should be noted that reconstruction based on Eq. (6) may produce unphysical results, i.e., the reconstructed density matrix is not positive semidefinite. In such a case, one may use several techniques to find a corresponding physical realization of the matrix. Simplest approach is to set negative eigenvalues to zero and renormalise the density matrix. More elaborated techniques find the closest positive semidefinite matrix using different norms using, for example, the maximum likelihood method [3, 4]. In the presented work, we use this method assuming that a distance between matrices is based on the Euclidean norm

$$d^2(\mathbb{A}, \mathbb{B}) = \sum_{ij} |\mathbb{A} - \mathbb{B}|^2_{ij}. \qquad (7)$$

Implementation of the technique allows us to obtain physical density matrix that best corresponds to the measured observables.

## 2. Reconstruction of the qutrit

While the discussion presented above is general, we now focus on a three-level system (qutrit), which is implemented in this work using the ground-state hyperfine level of $^{87}$Rb of the total angular momentum $f = 1$ (Fig.1). In this case, the density matrix can be written as

$$\rho = \begin{bmatrix} \rho^R_{\bar{1}\bar{1}} & \rho^R_{\bar{1}0} + i\rho^I_{\bar{1}0} & \rho^R_{\bar{1}1} + i\rho^I_{\bar{1}1} \\ \rho^R_{\bar{1}0} - i\rho^I_{\bar{1}0} & 1 - \rho^R_{\bar{1}\bar{1}} - \rho^R_{11} & \rho^R_{01} + i\rho^I_{01} \\ \rho^R_{\bar{1}1} - i\rho^I_{\bar{1}1} & \rho^R_{01} - i\rho^I_{01} & \rho^R_{11} \end{bmatrix}, \qquad (8)$$

where $\bar{1} = -1$. Additionally we assume the normalization condition $\text{Tr}(\rho) = 1$.[1] It can be shown [1] that, in a system probed on the $f = 1 \rightarrow F = 2$ transition (Fig.1), the observables $\hat{\alpha}_{R,I}$ and $\hat{\beta}$ take the forms:

$$\hat{\alpha}_R = \frac{\hat{d}_+^2 + \hat{d}_-^2}{\|\hat{d}\|} = \frac{1}{30}\left(|\bar{1}\rangle\langle 1| + |1\rangle\langle\bar{1}|\right), \tag{9a}$$

$$\hat{\alpha}_I = \frac{i\hat{d}_+^2 - i\hat{d}_-^2}{\|\hat{d}\|} = \frac{1}{30}\left(i|\bar{1}\rangle\langle 1| - i|1\rangle\langle\bar{1}|\right), \tag{9b}$$

$$\hat{\beta} = \frac{\hat{d}_+\hat{d}_- - \hat{d}_-\hat{d}_+}{\|\hat{d}\|} = \frac{1}{6}\left(|\bar{1}\rangle\langle\bar{1}| - |1\rangle\langle 1|\right), \tag{9c}$$

where $\hat{d}_\pm$ are the dipole operators associated with the right- (+) and left-handed (−) circular polarization components of the linearly polarized probe light, and $\|d\|$ is the reduced dipole-matrix element corresponding to the transition.

To expand the set of observables used for tomography, we introduce the control pulses $U_i$ implemented as pulses of a DC/non-oscillating magnetic field that induce a geometrical rotation of the spin polarization by $\pi/2$ around the **x**- or **y**-axes. The explicit forms of the control pulses are

$$U_x = \frac{1}{2}\begin{bmatrix} 1 & i\sqrt{2} & -1 \\ i\sqrt{2} & 0 & i\sqrt{2} \\ -1 & i\sqrt{2} & 1 \end{bmatrix}, \tag{10a}$$

$$U_y = \frac{1}{2}\begin{bmatrix} 1 & -\sqrt{2} & 1 \\ \sqrt{2} & 0 & -\sqrt{2} \\ 1 & \sqrt{2} & 1 \end{bmatrix}. \tag{10b}$$

Applying $U_x$ and $U_y$ to the operators given by Eqs. (9) allows one to relate the observables $\alpha_{R,I}$ and $\beta$ with other density-matrix element and hence reconstruct the whole density matrix.

### 3. Derivation of scaling factors

#### 3.1. Global Scaling Factor

An essential parameter of the presented QST technique is the global scaling factor $\eta(\Delta)$ [see Eq. (2)]. The most convenient way to calculate this parameter is based on the measurements of the absorption of light in a completely unpolarized atoms. From the formula describing the change in the amplitude of light traversing a medium given in Ref. [1], one obtains

$$\frac{\delta E}{E}(t;\Delta) = \frac{2\eta(\Delta)}{2f+1}\sum_{n=-f}^{f}\begin{pmatrix} f & 1 & F \\ -n & 1 & n-1 \end{pmatrix}^2, \tag{11}$$

where $f$ and $F$ are the total angular momentum of the ground and excited states, respectively, which for our system are $f = 1$ and $F = 2$. Additionally, one should notice that this model was derived for a two–level system, while in more complex systems of $^{87}$Rb, where the two ground states $f = 1$ and $f = 2$ exist, the absorption of light tuned to the $f = 1$ ground state results only

---

[1]It is noteworthy that the normalisation of the density matrix modifies the vector **b** by shifting its elements by a fixed value.

from 3/8 of the total number of atoms in the sample. As in our experimental procedure, we repump all the atoms in the $f = 1$ state, Eq. (11) needs to be modified to account for the fact

$$\frac{\delta E}{E}(t;\Delta) = \frac{8}{3}\frac{2\eta(\Delta)}{3}\sum_{n=-1}^{1}\begin{pmatrix} 1 & 1 & 2 \\ -n & 1 & n-1 \end{pmatrix}^2 = \frac{16}{27}\eta(\Delta). \tag{12}$$

Since the amplitude of light is not a parameter easily measurable experimentally, it is more convenient to replace a light-amplitude change by the light-intensity change

$$\frac{\delta E}{E} = \sqrt{1+\frac{\delta I}{I}} - 1, \tag{13}$$

where $I$ is the probing light intensity and $\delta I$ is the change of the light intensity resulting from the absorption in the medium (note that $\delta I$ is negative).

In a real experiment, the intensity of transmitted light can be influenced by losses at different optical elements within the system (e.g., glass walls of the vapor cell). For a reliable reconstruction, it is crucial to differentiate these losses from the light absorption in the medium. To address this, we measure the sum of the two outputs of a balanced polarimeter used for the state reconstruction and the intensity of the light before its transition through the atomic ensemble (Fig. 2). This

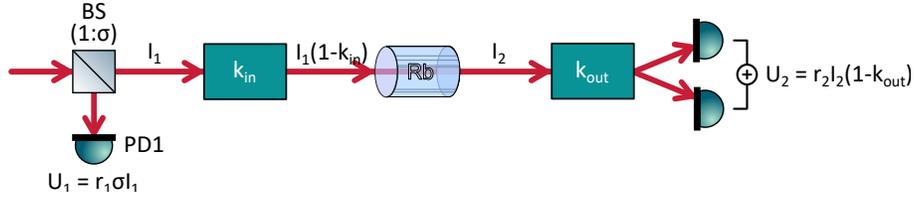

Fig. 2. Simplified scheme of the setup used for the measurements of the global scaling factor $\eta(\Delta)$. A non-polarization beam splitter (BS) of the 1:$\sigma$ splitting ratio is used to measure the light intensity before the atomic ensemble. The light intensity is measured as a voltage $U_1(\Delta) = r_1\sigma I_1(\Delta)$, where $r_1$ is the light power-voltage conversion factor in a transimpedance photodetector detector (PD1). The light intensity is written as a function of the light detuning $\Delta$ to remark the fact that changes in the light frequency usually introduce changes in the light intensity emitted by the laser. After BS, there are various optical elements, introducing intensity loses $k_{\text{in}}$, and a vapor cell containing atoms whose state one wants to reconstruct. Next, the beam experiences losses $k_{\text{out}}$ due to the elements situated after the cell and finally the light is split using the balanced polarimeter, consisting of a Wollastone prism and two PDs. The output signal $U_2$ is the sum of the voltages measured by each of the PDs of the same conversion gain $r_2$, i.e., $U_2(\Delta) = r_2 I_2(\Delta)(1-k_{\text{out}})$.

allows us to write

$$\frac{\delta I}{I} = \frac{I_2(\Delta) - I_1(\Delta)(1-k_{\text{in}})}{I_1(\Delta)(1-k_{\text{in}})} = \frac{r_1\sigma}{r_2(1-k_{\text{in}})(1-k_{\text{out}})}\frac{U_2(\Delta)}{U_1(\Delta)} - 1. \tag{14}$$

The most convenient way to determine the overall scaling [corresponding to the first term in Eq. (14)] is to measure the signal for far-detuned light $\Delta \to +\infty$. As in such a case absorption is negligibly small, one obtains

$$\frac{U_1(+\infty)}{U_2(+\infty)} = \frac{r_1\sigma}{r_2(1-k_{\text{in}})(1-k_{\text{out}})}. \tag{15}$$

Combining Eqs. (12)–(15), we obtain the final relation for the global scaling factor $\eta(\Delta)$, which is given by

$$\eta(\Delta) = \frac{27}{16}\left(\sqrt{\frac{U_1(+\infty)}{U_2(+\infty)}\frac{U_2(\Delta)}{U_1(\Delta)}} - 1\right). \tag{16}$$

## 3.2. Local Scaling Factor

The second important parameter that is needed for the tomography is the local scaling factor $\zeta(\Delta)$. As discussed in the main text, to calculate the parameter we compare the rotation signals measured for stretched states pumped along the **x**- and **z**-axes, which are given by the density matrices

$$\rho_x = (1-\epsilon)\begin{bmatrix} \frac{1}{4} & -\frac{1}{2\sqrt{2}} & \frac{1}{4} \\ -\frac{1}{2\sqrt{2}} & \frac{1}{2} & -\frac{1}{2\sqrt{2}} \\ \frac{1}{4} & -\frac{1}{2\sqrt{2}} & \frac{1}{4} \end{bmatrix} + \frac{\epsilon}{3}\mathbb{1}, \tag{17a}$$

$$\rho_z = (1-\epsilon)\begin{bmatrix} 0 & 0 & 0 \\ 0 & 0 & 0 \\ 0 & 0 & 1 \end{bmatrix} + \frac{\epsilon}{3}\mathbb{1}. \tag{17b}$$

Here we assumed a small imperfection in pumping manifesting through the isotropic term of the amplitude $\epsilon$. It is noteworthy that none of the observables used for the state reconstruction [given by Eqs. (9)] generates any signal corresponding to the isotropic term ($\epsilon = 0$). Using Eqs. (9) in conjunction with Eq. (2) allows one to calculate the corresponding rotation signals for the two cases

$$\delta\alpha^{(z)}(t;\Delta) = -\frac{5(1-\epsilon)}{24}\eta(\Delta)\zeta(\Delta)e^{-\gamma_2 t}, \tag{18a}$$

$$\delta\alpha^{(x)}(t;\Delta) = -\frac{(1-\epsilon)}{48}\eta(\Delta)e^{-\gamma_1 t}\cos(2\Omega_L t). \tag{18b}$$

By including the amplitudes of the oscillatory and DC parts of the signals, the local scaling factor can be calculated as

$$\zeta(\Delta) = \frac{1}{10}\frac{\delta\alpha^{(z)}(0;\Delta)}{\delta\alpha^{(x)}(0;\Delta)}. \tag{19}$$

## 3.3. CYCLOPS-like measurements

In a real experiment, an oscillating part of the polarization-rotation signal may experience additional phase shift $\varphi$ and the whole signal may be DC offseted $D$. Thus, Eq. (2) is modified to

$$\delta\alpha(t;\Delta) = \eta(\Delta)\left([\langle\hat{\alpha}_R\rangle\sin(2\Omega_L t + \varphi) + \langle\hat{\alpha}_I\rangle\cos(2\Omega_L t + \varphi)]e^{-\gamma_1 t} - \zeta(\Delta)\langle\hat{\beta}\rangle e^{-\gamma_2 t}\right) + D. \tag{20}$$

These additional factors may arise due to electronics or the imbalance of the polarimeter. As the reconstruction method requires fine determination of the quadrature components of the signal, one conclude that knowing the precise value of $\varphi$ is needed. In our work, we use a slightly

different approach, incorporating a $\pi$-rotation around the **y**-axis, i.e.,

$$U_y(\alpha) = \begin{bmatrix} \cos^2(\frac{\alpha}{2}) & -\frac{\sin(\alpha)}{\sqrt{2}} & \sin^2(\frac{\alpha}{2}) \\ \frac{\sin(\alpha)}{\sqrt{2}} & \cos(\alpha) & -\frac{\sin(\alpha)}{\sqrt{2}} \\ \sin^2(\frac{\alpha}{2}) & \frac{\sin(\alpha)}{\sqrt{2}} & \cos^2(\frac{\alpha}{2}) \end{bmatrix} \implies U_y(\pi) = \begin{bmatrix} 0 & 0 & 1 \\ 0 & -1 & 0 \\ 1 & 0 & 0 \end{bmatrix}. \quad (21)$$

This results in a sign change in $\langle \hat{\alpha}_I \rangle$ and $\langle \hat{\beta} \rangle$

$$\langle \hat{\alpha}_R^{(Y)} \rangle = \langle U_y(\pi)^\dagger \hat{\alpha}_R U_y(\pi) \rangle = \langle \hat{\alpha}_R \rangle, \quad (22a)$$

$$\langle \hat{\alpha}_I^{(Y)} \rangle = \langle U_y(\pi)^\dagger \hat{\alpha}_I U_y(\pi) \rangle = -\langle \hat{\alpha}_I \rangle, \quad (22b)$$

$$\langle \hat{\beta}^{(Y)} \rangle = \langle U_y(\pi)^\dagger \hat{\beta} U_y(\pi) \rangle = -\langle \hat{\beta} \rangle. \quad (22c)$$

For such a pulse, the rotation signal is given by

$$\delta\alpha^{(Y)}(t;\Delta) = \eta(\Delta) \left\{ e^{-\gamma_1 t} \left[ \langle \hat{\alpha}_R \rangle \sin(2\Omega_L t + \varphi) - \langle \hat{\alpha}_I \rangle \cos(2\Omega_L t + \varphi) \right] + \zeta(\Delta) e^{-\gamma_2 t} \langle \hat{\beta} \rangle \right\} + D. \quad (23)$$

Calculating the difference between the signal with and without the pulse, one obtains

$$\left(\delta\alpha - \delta\alpha^{(Y)}\right)(t;\Delta) = 2\eta(\Delta) \left[ -\zeta(\Delta) e^{\gamma_2 t} \langle \hat{\beta} \rangle + e^{-\gamma_1 t} \langle \hat{\alpha}_I \rangle \cos(2\Omega_L t + \varphi) \right]. \quad (24)$$

As seen, this difference depends only on a single oscillatory component, thus there is no problem with determining the initial phase. Additionally, one sees that the difference is free from the offset due to the imbalance of the polarimeter.

In a similar fashion, one can obtain a signal containing the second quadrature using the following sequence of two pulses: the first pulse introducing the rotation around the **y**-axis by $\pi$ (being identical to the previous pulse), and the second pulse generating the rotation by $\pi/2$ around the **z**-axis

$$U_z(\alpha) = \begin{bmatrix} e^{-i\alpha} & 0 & 0 \\ 0 & 1 & 0 \\ 0 & 0 & e^{i\alpha} \end{bmatrix} \implies U_{zy} = U_z(\pi/2) U_y(\pi) = \begin{bmatrix} 0 & 0 & -i \\ 0 & -1 & 0 \\ i & 0 & 0 \end{bmatrix}. \quad (25)$$

The pulses transform the expectation values of the observables as

$$\langle \hat{\alpha}_R^{(ZY)} \rangle = \langle U_{zy}^\dagger \hat{\alpha}_R U_{zy} \rangle = -\langle \hat{\alpha}_R \rangle, \quad (26a)$$

$$\langle \hat{\alpha}_I^{(ZY)} \rangle = \langle U_{zy}^\dagger \hat{\alpha}_I U_{zy} \rangle = \langle \hat{\alpha}_I \rangle, \quad (26b)$$

$$\langle \hat{\beta}^{(ZY)} \rangle = \langle U_{zy}^\dagger \hat{\beta} U_{zy} \rangle = -\langle \hat{\beta} \rangle, \quad (26c)$$

which allows one to calculate the rotation difference

$$\left(\delta\alpha - \delta\alpha^{(ZY)}\right)(t;\Delta) = 2\eta(\Delta) \left[ -\zeta(\Delta) e^{\gamma_2 t} \langle \hat{\beta} \rangle + e^{-\gamma_1 t} \langle \hat{\alpha}_R \rangle \sin(2\Omega_L t + \varphi) \right]. \quad (27)$$

Similarly, to the former case, the signal only contains one quadrature and no dependence on the polarimeter imbalance.

The robustness of the signal against the unknown phase shift or signal offset motivated us to used the CYCLOPS pulses for our quantum-state reconstruction.

## 4. Example of the experimental reconstruction

In order to ullustrate reconstruction of a qutrit state, we utilize six experimental polarization-rotation presented in Fig. 3. The signals were observed for the state generated with a circularly polarized light propagating along the **x**-axis.

The reconstruction process involved several stages. First, we measured absorption of the probing light in an unpolarized sample to determine the global scaling factor (Sec. 3.1). In the considered experiment, we got $\eta(\Delta) = 0.3236(21)$. To determine the local scaling factor, we pumped our ensemble using a circularly polarized light propagating along the **x**-axis to prepare a stretched state in the direction (Sec. 3.2). We compare the signal with the similar results obtained with and without a magnetic $\pi/2$-pulse applied along the **y**-direction. This allowed us to determine the local scaling factor, which, in the considered case, was $\zeta(\Delta) = -0.4790(12)$.

Next, we measure nine signals. For each control-pulse sequence (no pulse, a $\pi/2$ rotation in the **y**-direction, and $\pi/2$ rotation around the **x**-direction), we employed three different sets of CYCLOPS pulses (no CYCLOPS pulse, a pulse in the **y**-direction, and a CYCLOPS pulse in the **z**- and **y**-directions). These signals are used to generate the difference signals [Eqs. (24) and (27)], which are are shown in Fig. 3. These signals are fitted to reconstruct the expectation values of our operators of interest. This is done with shared parameters, including the frequency, global phase, and relaxation rates.

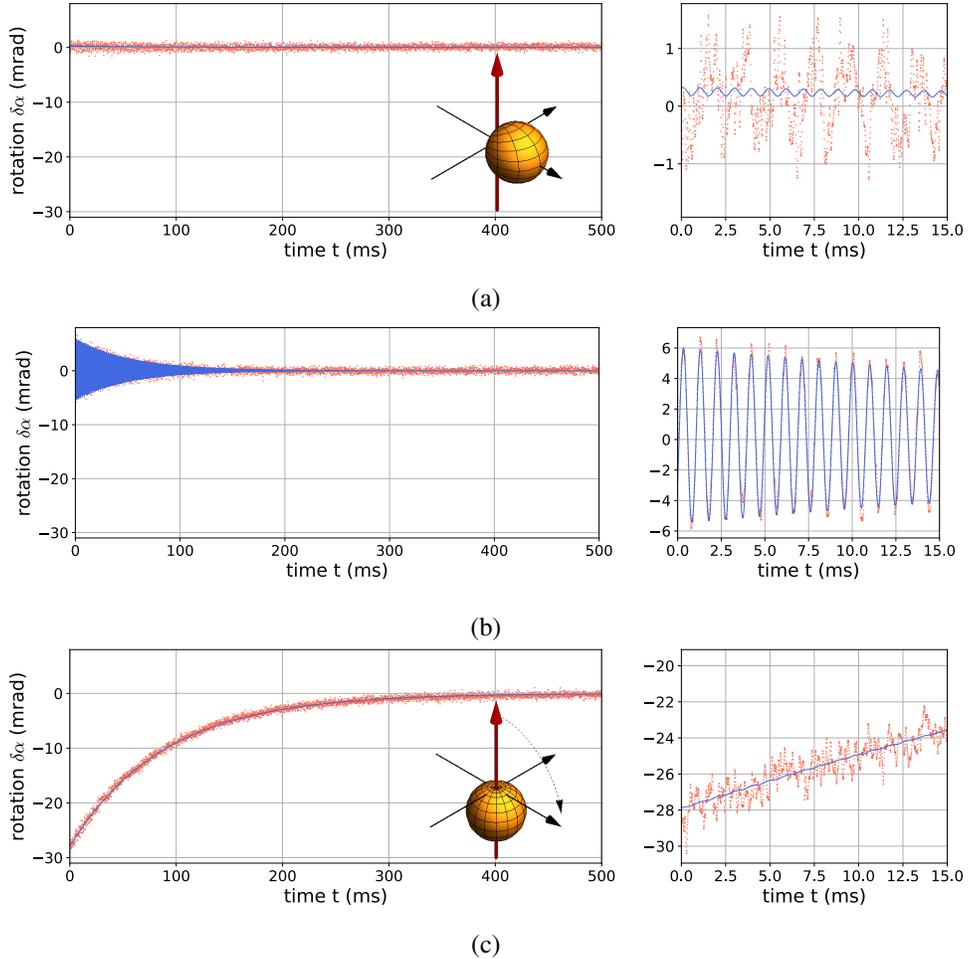

(a)

(b)

(c)

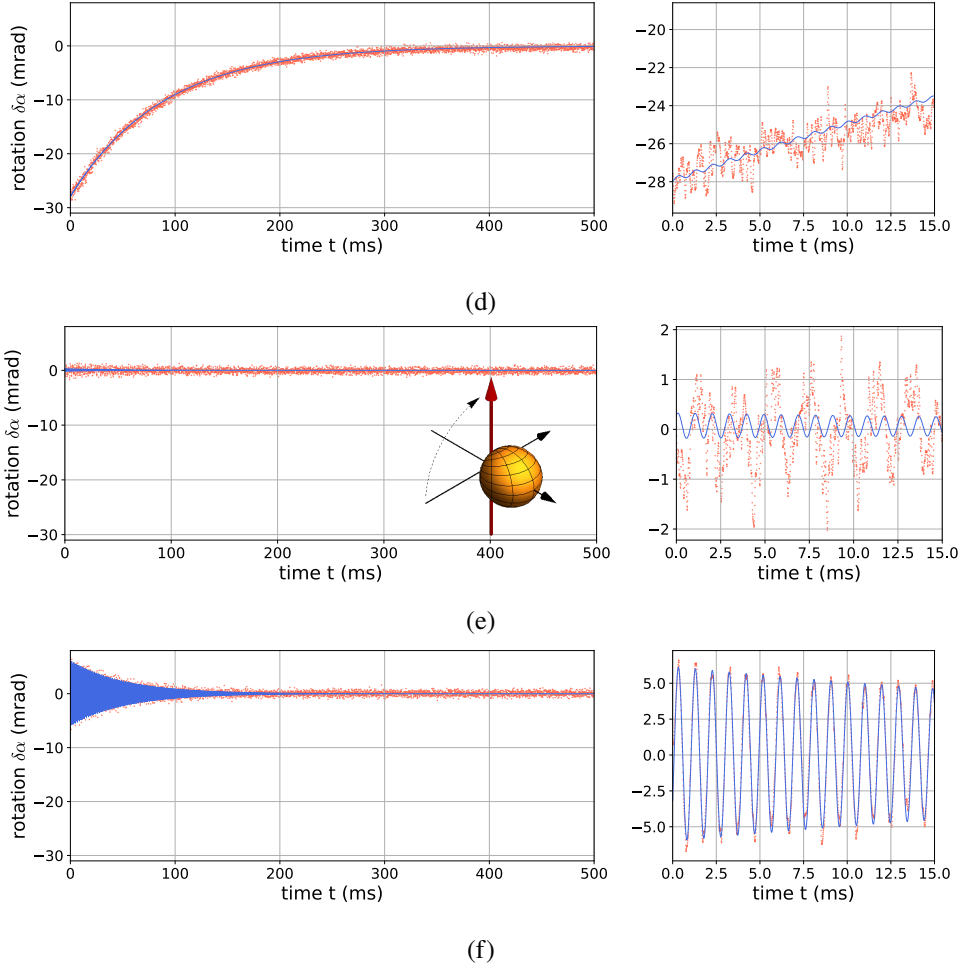

Fig. 3. Detected polarization rotation (red dots) and fitted dependences (blue lines), corresponding to the first [Eq. (24)] (a) and second [Eq. (27)] (b) quadrature of the signal measured without the control pulses. (c) and (d) correspond to the signals measured with the pulse rotating the state by $\pi/2$ around the **y**-axis and (e) and (f) with the pulse rotating the state by $\pi/2$ around the **x**-axis. The right column shows magnifications of the first 15 ms of the registered signals. The insets in (a), (c), and (e) show the angular-momentum probability surface [5] of the reconstructed state after respective rotations (denoted with black dashed arrows). The red axis denotes the axis of the magnetic field applied during detection, coinciding with the propagation direction. The graphical representation of the density matrix introduces intuition about the symmetry of the generated state.

Next, we use nine reconstructed observables to perform linear inversion [according to Eq. (6)] to obtain the matrix corresponding to the state. In the considered case, we obtained

$$\rho = \begin{bmatrix} 0.2363 & -0.3903 - 0.0002i & 0.2748 - 0.0033i \\ -0.3903 + 0.0002i & 0.5175 & -0.3737 + 0.0022i \\ 0.2748 + 0.0033i & -0.3737 - 0.0022i & 0.2461 \end{bmatrix}.$$

It is evident that the reconstructed matrix is not positive semidefinite, as indicated by its

eigenvalues eig($\rho$) = {1.057, −0.041, −0.016}. In order to address this problem, as a final step, we applied the maximum likelihood method [4] to find the closest positive semidefinite matrix to the reconstructed one. This process allowed us to reconstruct the following density matrix

$$\rho_{ML} = \begin{bmatrix} 0.2410 & -0.3507 + 0.0003i & 0.2447 - 0.0020i \\ -0.3507 - 0.0003i & 0.5104 & -0.3562 + 0.0027i \\ 0.2447 + 0.0020i & -0.3562 - 0.0027i & 0.2486 \end{bmatrix},$$

with nonegative eigenvalues eig($\rho$) = {1., 0., 0.} This state is also presented graphically in the main text in Fig. 2(a).



\end{document}